\documentclass[useAMS,usenatbib]{mn2e}

\usepackage{siunitx}
\usepackage[fleqn]{amsmath}
\usepackage{graphicx}
\usepackage{xcolor}
\newcommand{\degree}{deg}

\title[Formation of the G-ring arc]{Formation of the G-ring arc}

\author[N. C. S. Araujo; E. Vieira Neto and D. W. Foryta]{%
  N. C. S. Araujo$^{1}$\thanks{E-mail: nilcasr@gmail.com (NCSA)};
  E. Vieira Neto$^{1}$\thanks{E-mail:  ernesto@feg.unesp.br (EVN)} and
  D. W. Foryta$^{2}$\thanks{E-mail: foryta@fisica.ufpr.br (DWF)} \\ 
  $^{1}$ Departamento de Matem\'{a}tica, Univ. Estadual Paulista, 12516-410 Guaratinguet\'{a}, 
  S\~{a}o Paulo, Brazil \\
  $^{2}$ Departamento de F\'{\i}sica, Univ. Federal do Paran\'{a}, 83531-940 Curitiba, 
  Paran\'{a}, Brazil}

\begin{document}

\date{Original form 2014 December 15: last revision 02 May 2016}

\pagerange{\pageref{firstpage}--\pageref{lastpage}} \pubyear{2016}

\maketitle

\label{firstpage}

\begin{abstract}
  
  Since 2004, the images obtained by Cassini spacecraft's on-board cameras have
  revealed the existence of several small satellites in the Saturn system. Some
  of these small satellites are embedded in arcs of particles. While these
  satellites and their arcs are known to be in corotation resonances with Mimas,
  their origin remains unknown. This work investigates one possible process for
  capturing bodies into a corotation resonance, which involves raising the
  eccentricity of a perturbing body. Therefore, through numerical simulations
  and analytical studies, we show a scenario that the excitation of Mimas'
  eccentricity could capture particles in a corotation
  resonance and given a possible explanation for the origin for the arcs.
  
\end{abstract}


\section{Introduction}

Since 2004, the Cassini spacecraft has returned to Earth a copious amount of
data about the Saturnian system.  In particular, its imaging system revealed the
existence of several small satellites. There are papers
\citep{Spitale.etal-2006,Cooper.etal-2008,Hedman.etal-2010a} showing that Mimas
perturbs the orbits of some of these satellites by resonant interactions.

In 2006 data returned by Cassini revealed an arc of dust inside the G-ring. Since then, this region has
been explored by \citet{Hedman.etal-2007} precisely calculated the G arc's mean motion
using the Cassini's data. It revealed that this arc is in a 7:6 corotation resonance with Mimas. 
Through the analysis of images obtained between 2008 and 2009,
it was found that the satellite Aegaeon was located inside this
arc \citep{Hedman.etal-2010a}. And since this satellite is
immersed in this arc, it is also trapped in this corotation resonance with
Mimas.

\citet{Hedman.etal-2010a} confirm this corotation resonance
through numerical simulations of the full equations of motion. They also
identify that this satellite is in corotation resonance with characteristic
angle
\begin{equation}
  \varphi_{{}_{CER}}=7\lambda_{{}_{Mimas}}-6\lambda_{{}_{Aegaeon}}-\varpi_{{}_{Mimas}}.
\end{equation}
Their simulations also show that a nearby Lindblad resonance also influences the
moon's motion \citep{Hedman.etal-2010a}.

Prior this study and also from Cassini images
two others satellites were discovered,
Methone in 2004 \citep{Spitale.etal-2006} and Anthe in 2007
\citep{Cooper.etal-2008}. The numerical analysis of the full equations of motion
\citep{Spitale.etal-2006, Cooper.etal-2008} indicates that these satellites are
in corotation resonances whose characteristic angles are, respectively,
\begin{alignat}{4}
  &
  \varphi_{{}_{CER}}&=15\lambda_{{}_{Methone}}&-14\lambda_{{}_{Mimas}}&-\varpi_{{}_{Mimas}},\\
  &
  \varphi_{{}_{CER}}&=11\lambda_{{}_{Anthe}}  &-10\lambda_{{}_{Mimas}}&-\varpi_{{}_{Mimas}}.
  &
\end{alignat}

The Cassini images also show that Aegaeon, Anthe and Methone are immersed in
tenuous arcs of particles \citep{Hedman.etal-2007, Hedman.etal-2009}
  and \citet{Hedman.etal-2010a} said that these particles probably have been
originated from the material knocked off, at low speeds, from the surface of
these satellites. Therefore, these particles don't have enough energy to escape the corotation
resonance and then remain close to the satellite filling the nearby
space. With these evidences \citet{Hedman.etal-2010a} concluded that the study
of these satellites and their arcs may improve the understanding of the
connection between satellites and rings. Because these objects and their arcs
are in corotation resonances, these satellites may be seen as a distinct class
of objects in the Saturn system.

The existence of these satellites immersed in arcs of dust
allows us to infer two possibilities for their origin. (i) The satellites were
build-up by particles previously caught in resonance and their arcs are the
vestiges of this formation.  Alternatively, (ii) Mimas had
captured satellites already formed, and consequently the arcs
had originated from particles that had broken
off from these satellites.

Although we are not able to judge immediately which of the above statements are
the correct one, we realize that both statements depend on the possibility of bodies been captured in corotation
resonance with Mimas, either small particles or a full satellite. Thus, in this
work we investigate the mechanism of capturing particles in corotation
resonance.

The corotation resonance exists only when the perturbing satellite has
eccentricity different from zero \citep{Murray.Dermott-1999}. Then, in our
problem, corotation depends on the eccentricity of Mimas. And it is known that, due to
tidal effects, the eccentricity of Mimas should be lower than the current
value \citep{Meyer.Wisdom-2008}. But Mimas has a higher
eccentricity than most of Saturn's regular satellites, it is likely that his
eccentricity was increased through a resonant interaction with another
satellite. In this work we will consider the presence of Enceladus playing an
important role in this scenario.

The aim of this paper is to investigate the mechanisms which would make Mimas
capture particles in a corotation resonance in the past of the Saturn system. The
study will be made through numerical simulations and
analytical studies which will be developed using the dynamics of three and four
bodies problems considering the effects of non-spherical shape of Saturn.

In our problem the main bodies are Saturn, Mimas, Enceladus and particles of
G-ring. We will study the scenario where particles will be captured in
corotation resonance with Mimas when Mimas passes through a resonance with
Enceladus. As a result of this study, we will have a better understanding of the
dynamics involved in the origin and stability of small
satellites. In this study, we will focus on 7:6 corotation
resonance

In section \ref{arepo} we make a brief discussion on the corotation
resonance. Section \ref{model} introduces the mechanism we
develop to obtain the eccentricity variation of
Mimas. Section \ref{memo} shows the effects of this mechanism on Mimas
orbit. Section \ref{captura} presents the results of this mechanism when a ring
of particles is immersed in the 7:6 corotation resonance region. In Section
\ref{evolution} we analyze the process of capture due the migration. Finally,
the concluding remarks for this study will be found in the section
\ref{conclusao}.

\section{Resonances in oblate planets}
\label{arepo}

When an object orbits an oblate planet, its orbit experiences effects of a
potential which depends on the planet's zonal harmonic coefficients $J_2$,
$J_4$, $\cdots$. The perturbations of that potential cause the rotation of the
orbit in space, the precession of the unperturbed orbit.

The rotation of the orbit generates three frequencies: $n$, the mean motion;
$\kappa$ and $\nu$, radial/vertical epicyclic frequency, respectively
\citep{Murray.Dermott-1999}. Thereby, considering the additional gravitational
effects of a perturbing satellite on a particle, when this system is around an
oblate planet, the orbit of the particle can be analysed through those
frequencies.

Those frequencies are associated with the precession of the node and pericenter
of the orbit by $\kappa = n - \dot{\varpi}$ and $\nu = n - \dot{\Omega}$ and
then the definition of the corotation resonance \citep{Murray.Dermott-1999}
occurs when:
\begin{equation}
  \varphi_{{}_{CR}} = j\lambda' + (k+p-j)\lambda - k\varpi' - p\Omega',
\end{equation}
where the primed orbital elements belong to the perturbing satellite while the non-primed
orbital elements belong to the particle, $j, k, p$ are
integer values.

When a particle is in corotation resonance with a perturbing satellite, some
orbital parameters are modified. There are analytical models able to estimate
the extent of these variations, for example, the Pendulum Model or the
Hamiltonian Approach \citep{Murray.Dermott-1999}. The orbital parameters that is
most strongly affected by a corotation resonance is the semi-major axis. Using
the Pendulum Model, we can calculate the maximum width libration of the
semi-major axis for corotation resonance.

The maximum width for a corotation resonance is (\citet{Murray.Dermott-1999},
Eq. (10.10))
\begin{equation}
  \label{larguradacorrotacao}
  W_{CR} = 8 \left( \frac{a|R|}{3Gm_p} \right)^{1/2} a,
\end{equation}
where $a$ is the semi-major axis of the perturbed body, $m_p$ is the central
body's mass and $R$ is the relevant term of the perturbing function, whose
equation is
\begin{equation}
  R = \frac{Gm'}{a'} f_d(\alpha) \, e'{}^{|k|} \, 
      s'{}^{|p|} \, \cos{\varphi_{{}_{CER}}},
\end{equation}
where the primed parameters are of the perturbing satellite, with $m'$, $a'$,
$e'$ and $s'$ as the mass, semi-major axis, eccentricity and a value associated
to the inclination $I'$, i.e. $s'=\sin(I'/2)$, $f_d(\alpha)$ is a function in
Laplace's coefficients for the directs terms of the perturbing function,
$\varphi_{{}_{CER}}$ is the corotation resonant angle, $k$ and $p$ are
integers.

Therefore, from the above equations, we expect that to existence of the
corotation resonance, the perturbing satellite's eccentricity must be different
from zero.

\section{Model}
\label{model}

The current eccentricity of Mimas is approximately 0.02,
and \citet{Meyer.Wisdom-2008} pointed out that this value is relatively high and
would imply a much higher value in the past, or it was recently excited. It is
known that the eccentricity of a satellite can be excited due to resonances
\citep{Murray.Dermott-1999}, but the recent value of Mimas' eccentricity cannot
be explained by present resonances such as Mimas-Tethys 4:2 mean motion
resonance \citep{Champenois.Vienne-1999, Callegari.Yokoyama-2010}. Thus we
suppose that Mimas experienced some event in the past that increased its
eccentricity.

\citet{Meyer.Wisdom-2008} suggested that Mimas was captured by Enceladus or by
Dione into a resonance while the eccentricity of Mimas was less than the current
value. They verified that when Mimas came into some
eccentricity-type resonance with one of these
satellites, Mimas' eccentricity increased and even exceed its current
value. After these satellites escaped from this resonance
interaction, the tidal orbital evolution decreased their eccentricity to the
current values, as we can notice in Figures 6, 7 and 9 from the paper of
\citet{Meyer.Wisdom-2008}. Therefore, these kind of resonant
encounters could have temporarily increased Mimas'
eccentricity, which have flavoured the capture of particles in Mimas' corotation
resonances.

In our study we adopt a scenario where we have Mimas-Enceladus 3:2 e-Mimas
resonance \footnote{e-Mimas resonance is the notation of
  \citet{Meyer.Wisdom-2008} for Mimas' eccentricity-type first order resonance.}
as discussed in \citet{Meyer.Wisdom-2008}. That resonance would induce an
increase in Mimas' eccentricity even larger than the current one, and after a
certain time those satellites could go out of that resonance. After the escape,
due to the tidal effects, the eccentricity of Mimas would decay to the current
value \citep{Meyer.Wisdom-2008}.

In our scenario Mimas and Enceladus were closer to Saturn than they are
today. Thus, we had to calculated a consistent position for them based on a
satellite tidal evolution. For this task, we perform the procedures discussed
below.

First, we evaluate the ratio of the
semi-major axes $\alpha$ when Mimas and Enceladus were trapped in the 3:2
eccentricity mean motion resonance. When two satellites are in mean motion
resonance, $\alpha$ remains approximately constant and can be calculated with
\citep{Champenois.Vienne-1999}
\begin{equation} 
  \label{alpha}
  \alpha = \left( \frac{p+q}{p} 
           \right)^{\!-2/3}  \!\!
           \left( 1 \! + \! \frac{ q_{{}_1} \dot{\varpi}_{{}_M} \! + \!
                             q_{{}_2} \dot{\varpi}_{{}_E} \! + \!
                             q_{{}_3} \dot{\Omega}_{{}_M} \! + \!
                             q_{{}_4} \dot{\Omega}_{{}_E}}{n_{{}_E}(p+q)}
           \right)^{\!-2/3}\hspace{-1.75em},
\end{equation}
where $p$, $q$, $q_{{}_1}$, $q_{{}_2}$, $q_{{}_3}$ and $q_{{}_4}$ are integers;
while $\dot{\varpi}_{{}_M}$, $\dot{\varpi}_{{}_E}$, $\dot{\Omega}_{{}_M}$ and
$\dot{\Omega}_{{}_E}$ are precession rates of longitude of pericenter, longitude
of ascending node for Mimas and Enceladus, respectively, and $n_{{}_E}$ is the
mean motion of Enceladus. Using the angles values for the 3:2 eccentricity mean
motion resonance in equation (\ref{alpha}), we get the ratio between the
semi-major axes of Mimas and Enceladus when they were trapped in that resonance
\citep{Champenois.Vienne-1999}. In this paper, the values of
$\dot{\varpi}_{{}_M}$, $\dot{\varpi}_{{}_E}$, $\dot{\Omega}_{{}_M}$ and
$\dot{\Omega}_{{}_E}$ are consistent with the geometric orbital elements
\citep{Renner.Sicardy-2006}. For the case of our resonance, we find the values
of $q_{{}_1}$, $q_{{}_2}$, $q_{{}_3}$ and $q_{{}_4}$ from the comparison between
general resonant angle,
\begin{equation} 
  \label{anguloressontegeral}
  \varphi = p        \lambda 
          - (p+q)    \lambda'
          + q_{{}_1} \varpi 
          + q_{{}_2} \varpi'
          + q_{{}_3} \Omega 
          + q_{{}_4} \Omega',
\end{equation}
where $\lambda$, $\varpi$, and $\Omega$ are the mean longitude, longitude of
pericenter and longitude of the ascending node, respectively for the inner
satellite, while those longitudes with prime represent the angles for the outer
satellite, with resonant angle as,
\begin{equation} 
  \label{emimasenceladus}
  \varphi_{e} = 2\lambda_{{}_M} - 3\lambda_{{}_E} + \varpi_{{}_M},
\end{equation}
where $\lambda_{M}$ is the mean longitude of Mimas, $\lambda_{E}$ is the mean
longitude of Enceladus. After our evaluations, we obtain
$\alpha$ equal to $0.7637895$.

The next step was to find the semi-major axis of the satellites corresponding to
our $\alpha$. We will call those semi-major axis ''ancient semi-major axes`` and
it can be evaluated through the equation (the development of
the following equation can be seen at appendix A)
\begin{equation}
  \label{semieixomaiormimasnopassado}
  a_{{}_{0M}} = \left[ \frac{a_{{}_M}^{13/2}\,\left(\dfrac{m_{{}_E}}{m_{{}_M}}\right)-a_{{}_E}^{13/2}}
                          {\left(\dfrac{m_{{}_E}}{m_{{}_M}}\right)-\dfrac{1}{\alpha^{13/2}}}
              \right]^{2/13},
\end{equation}
where $a_{{}_{0M}}$ is the Mimas' ancient semi-major axis, $a_{{}_M}$,
$a_{{}_E}$, $m_{{}_M}$ and $m_{{}_E}$ are the current semi-major axes and mass
of the satellites Mimas and Enceladus, respectively.

Equation (\ref{semieixomaiormimasnopassado}) is a function
of $\alpha$, the Mimas and Enceladus' semi-major axes ratio. As we had used a
value of $\alpha$ when both satellites were in resonance with each other, we can
locate the ancient semi-major axis of Enceladus using
$a_{{}_{0E}}={a_{{}_{0M}}/\alpha}$. The results of $a_{0_{E}}$ and $a_{{}_{0M}}$
indicate that those satellites were more distant from each other in the past
compared with their current positions. As expected, since the inner satellite
migrate faster than the outer one.

The values of Mimas and Enceladus ancient semi-major axis, calculated through equation
(\ref{semieixomaiormimasnopassado}), are
approximated values. To obtain values which lead the
system to enter in the resonance we try some values near the ancient semi-major
axes until we got them close to a position in the verge of resonance.

The tidal effect can cause Saturn's satellites migration with velocity orders
between of \SI{e-7}{} and \SI{e-5}{km} per year
at their current positions \citep{Meyer.Wisdom-2008,Lainey.etal-2012}. Therefore, we should migrate
Mimas and Enceladus with velocities close to those velocities. However, if we
did that, it would take a long computational time, as we have noted in migration
tests using that order of magnitude for the migration rate. As we are trying to prove the concept of corotation
eccentricity resonance capture caused by the eccentricity of
Mimas when it was enlarged due to mean motion resonance
with Enceladus, therefore we used values which make our
simulations more quickly. Thus, we migrate Mimas with a rate for the semi-major axis close to \SI{1}{km} per year and Enceladus around
\SI{0.01}{km} per year. The external satellite will migrate slower than the
internal one (see equation (\ref{taxasemieixomario1}) and also
\citealt{Burns.Matthews-1986}). To generate those rates, we inserted into the
dynamic of each satellite a drag force
\begin{equation} 
  \label{forcaperturbadora}
  \vec F_p = - \gamma v\vec v,
\end{equation}
where $\vec{v}$ is the velocity of the satellite, and $\gamma$ a constant which
will give the mentioned velocities for the semi-major axis. The value of
semi-major axis will increase due to a negative sign for $\gamma$.

\begin{figure}
\centerline{\includegraphics[width=\linewidth]{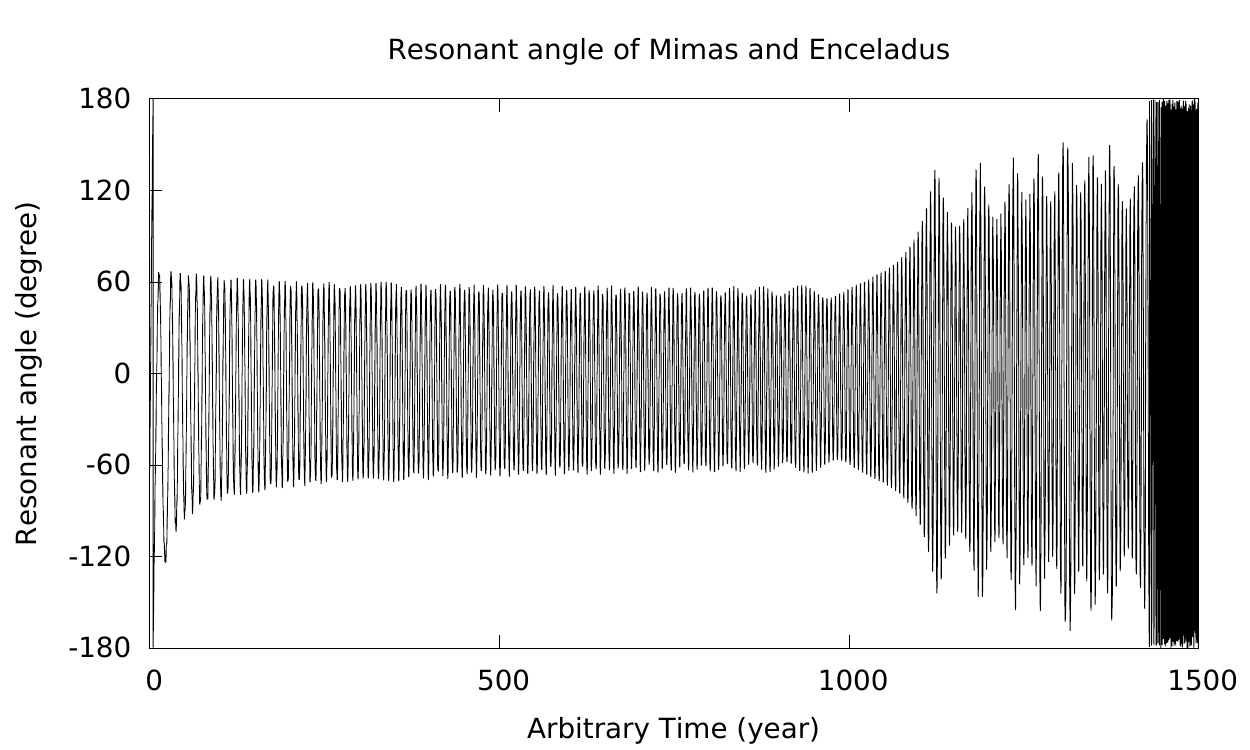}}
\caption{Critical angle for Mimas-Enceladus 3:2 e-Mimas resonance (equation
  (\ref{emimasenceladus})) during the migration process. It shows a libration
  around zero. To the end of tandem migration the critical angle increases its
  sweep following the increase on the eccentricity of Mimas.}
 \label{anguloressonantemimasenceladusmigrando}
\end{figure}

In a scenario with a lower eccentricity value for Enceladus, using the
  three-bodies dynamics for Saturn, Mimas and Enceladus, we found no chaotic
  behaviour for the resonance angle when we have Mimas-Enceladus 3:2 e-Mimas
  resonance. That is, we have found the same results that
\citet{Meyer.Wisdom-2008} had encountered for Mimas-Enceladus 3:2 e-Enceladus
resonance. It means that Mimas does not escape from the resonance capture during
our time integration. This integration had run for \SI{3000}{years} of arbitrary time (equivalent approximately to 300 million years, if
  we have used the corrects rates for migration).

Tidal effects could have reduced a higher eccentricity of
Enceladus, and in our scenario we verified that this higher eccentricity is
necessary for Mimas to escape from the resonance. Higher values for Enceladus's
eccentricity are possible due to earlier captures into others resonances as can
be seen in \citet{Meyer.Wisdom-2007}.

We used respectively \SI{0.005}{} and
  \SI{0.02}{} for the initial eccentricities of Mimas and Enceladus. These
values make our scenario works very well as we will show below. Probably they
may exist other mechanisms which could take Mimas out of the resonance other
than this value for the eccentricity of Enceladus,
but we are most interested in proving that the corotation resonance is able to
  capture particles, and we will investigate those features in other works.

\section{Migration Effects on Mimas Orbit}
\label{memo}

\begin{figure}
 \hbox to \linewidth{\includegraphics[width=\linewidth]{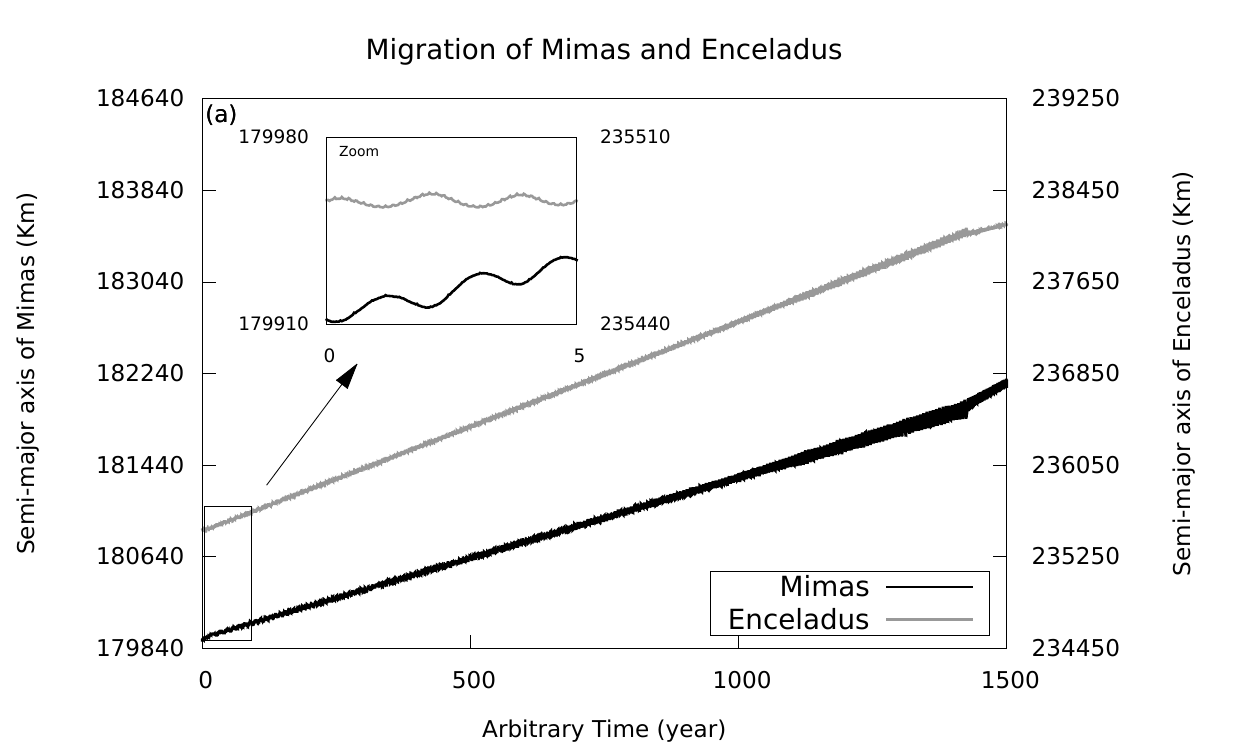}}
 \hbox to \linewidth{\hskip 1.1pc\includegraphics[width=0.781\linewidth]{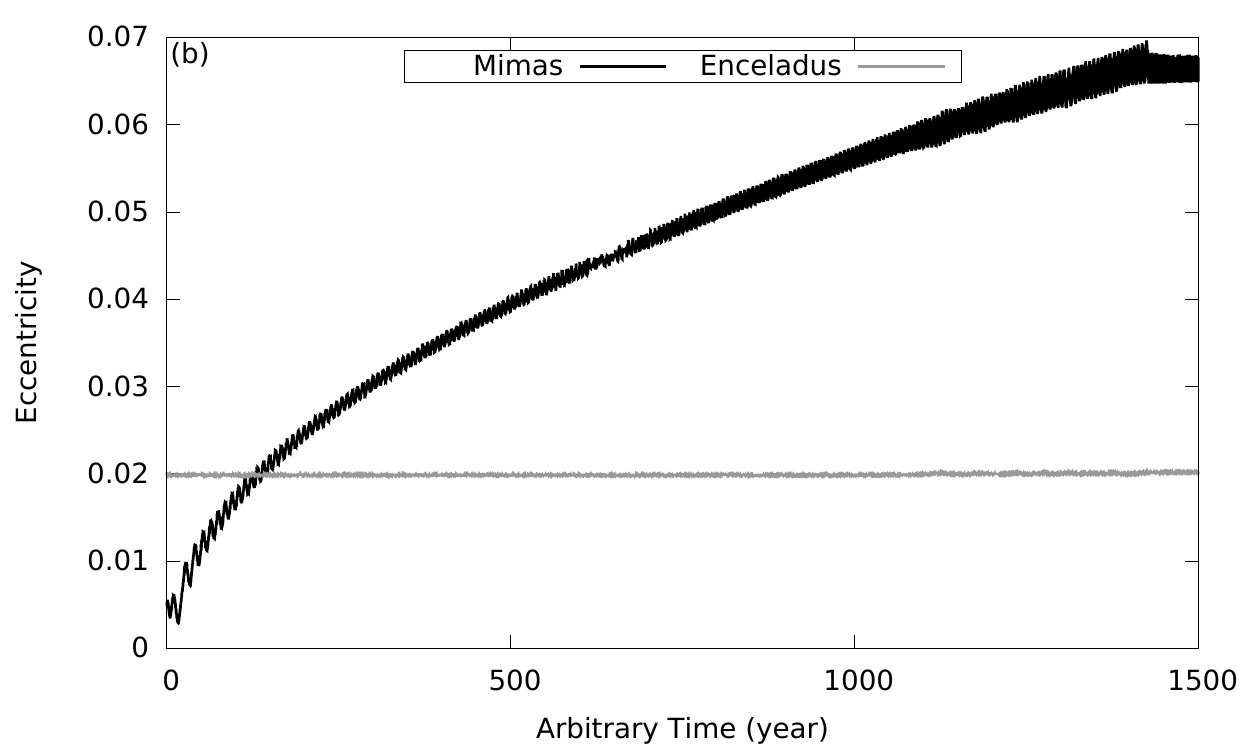}}
 \caption{When Mimas is caught in resonance by Enceladus its eccentricity grows
   as seen in the lower panel. During this time a tandem migration is established
   resulting in a net migration rate, that differs from the tidal evolution
   alone, as seen in upper plot. The net migration rate of Enceladus raises by
   the interaction with Mimas while Mimas' migration rate decreases. This effect
   ends after the escape of the resonance and the migration rates change, as can
   be seen in the end of the figure (a).}
 \label{semieixomaiormimasenceladusmigrando}
\end{figure}

\begin{figure*}
 \hbox to \linewidth{%
  \includegraphics[width=0.525\linewidth]{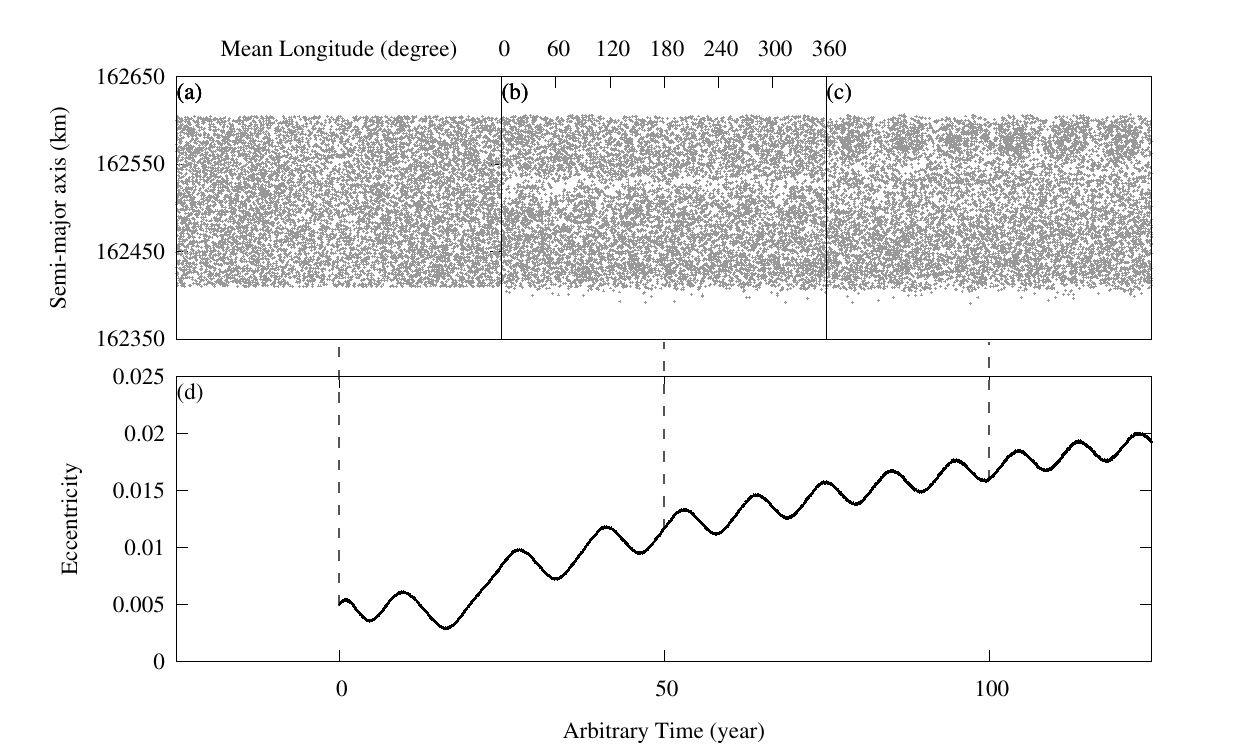}%
  \hss%
  \includegraphics[width=0.525\linewidth]{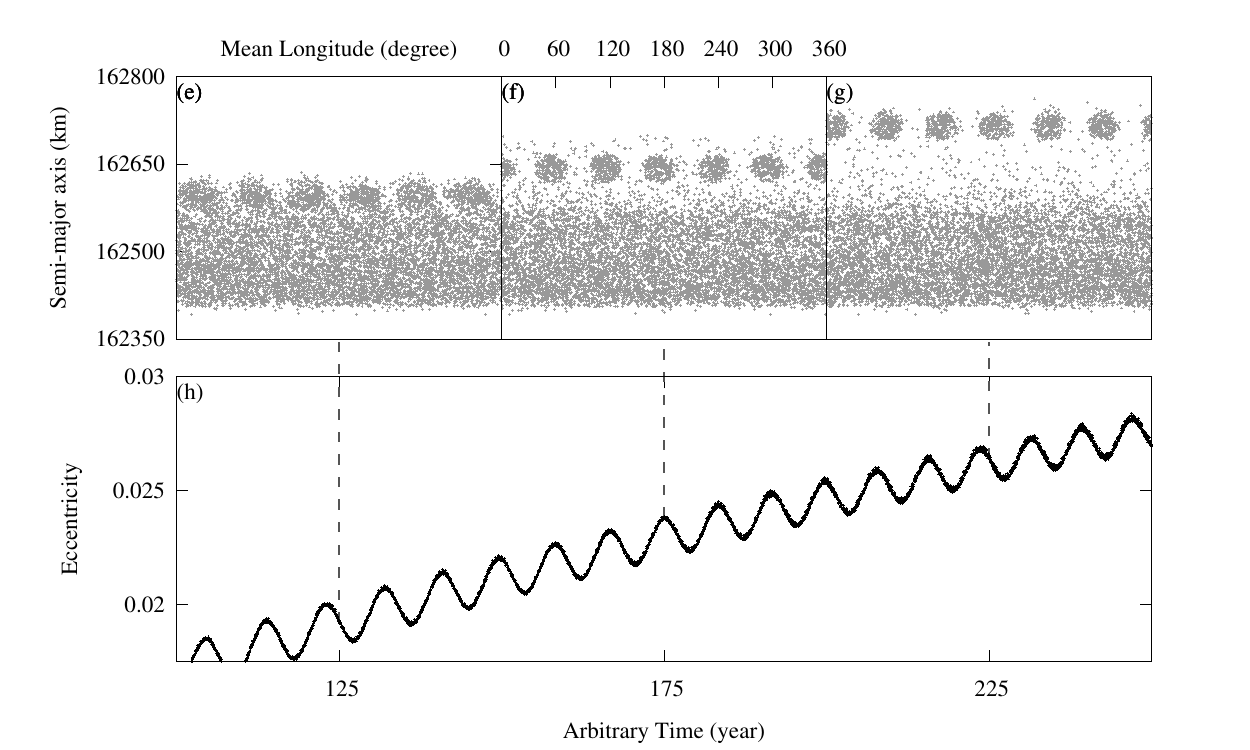}%
 }%
 \caption{These six snapshots show the evolution of the capture in corotation
   resonance. The simulation began when occurs the coupling migration between
   Mimas and Enceladus. In gray a full ring of particles representing possible
   bodies that may be captured inside the corotation resonance. In last snapshot,
   we can see all corotations sites which had been captured particles and had
   been conveyed outward the ring by the interaction with Mimas in its migration
   process.}
 \label{migracaomimasenceladusmA}
\end{figure*}

In the first part of our hypotheses, we treat the excitation on
Mimas eccentricity due to its passage through the
Mimas-Enceladus 3:2 e-Mimas resonance during the migration
  process of these satellites.  To perform this simulation we used the
Gauss-Radau spacings described by \citet{Everhart-1985} with initial time step
of 0.1 day. The dynamics of this process included Mimas,
Enceladus and an oblate Saturn, and also the drag force of equation
(\ref{forcaperturbadora}). For the initial conditions of Mimas and Enceladus, we
use the consideration commented in Section \ref{model} to form the set of
initial condition shown in table \ref{initialcondition}.

\begin{table}
  \caption{\label{initialcondition} Initial Conditions of Enceladus and Mimas in Orbital Elements, after changes commented in Section \ref{model}.%
  }
  \begin{center}
    \begin{tabular}{ll}
       \hline
      Name & Mimas   \\
      \hline
       Mass      & \SI{3.75E+22}{g}  \\     
       Radius    & \SI{198.8}{km} \\ 
       $a$       & \SI{1.799177e+05}{km} \\
       $e$       & \SI{5.e-03}{}\\
       $I$       & \SI{1.563223e+00}{deg} \\
       $\Omega$  & \SI{3.564454e+02}{deg} \\
       $\varpi$  & \SI{1.899002e+02}{deg} \\
       $\lambda$ & \SI{3.166168e+02}{deg} \\
      \hline
      Name & Enceladus   \\
      \hline
       Mass      & \SI{10.805E+22}{g}  \\     
       Radius    & \SI{252.3}{km}  \\
       $a$       & \SI{2.354949e+05}{km}   \\
       $e$       & \SI{2.e-02}{}   \\
       $I$       & \SI{5.067876e-03}{deg}\\
       $\Omega$  & \SI{2.633612e+02}{deg}  \\
       $\varpi$  & \SI{2.822281e+02}{deg} \\
       $\lambda$ & \SI{2.950096e+02}{deg} \\
        
      \hline
    \end{tabular}
  \end{center}
\end{table}

In Figure (\ref{anguloressonantemimasenceladusmigrando}) we observe the
behaviour of the critical angle for Mimas-Enceladus 3:2 e-Mimas resonance
(equation (\ref{emimasenceladus})). This behaviour occurs
  due to the passage of Mimas and Enceladus through the resonance during the
migration process. We can see that at the very beginning of the simulation the
Mimas-Enceladus system was not in resonance (the resonant angle circulates) and,
due to the system migration, those satellites enter in the Mimas-Enceladus 3:2
e-Mimas resonance \citep{Meyer.Wisdom-2008} in which the resonant angle librates
around \SI{0}{\degree}. We can see along
  the migration process a decreasing amplitude for the resonant angle, it may
  turn the resonance between them more robust. Subsequently, this amplitude
begins to increase until it reaches the value of
\SI{180}{\degree}. At this point, we can say that Mimas and
Enceladus are out of the resonance.

In Figure (\ref{semieixomaiormimasenceladusmigrando}a) we can see the behaviour
of Mimas' and Enceladus' semi-major axes during the
migration process. Before they enter in resonance, the variation rates for the semi-major are the
ones stated in the last section, with Enceladus migrating slower than Mimas (see
zoom in Figure
(\ref{semieixomaiormimasenceladusmigrando}a)). After they enter
in resonance, the rate of  Enceladus' semi-major axis
variation increases until reaching a value slightly larger than Mimas, while
the semi-major axis rate variation for Mimas decreases. These can be seen in the semi-major axis
inclination showed in Figure (\ref{semieixomaiormimasenceladusmigrando}a). When
they come out of resonance, the variation rates for the
semi-major axis returns to their previous values.

This passage through the resonance affects Mimas' eccentricity significantly, as
shown in Figure (\ref{semieixomaiormimasenceladusmigrando}b). When Mimas is in resonance, its
eccentricity increases to high values while Enceladus eccentricity remains
constant. Despite the migration velocity, we have adopted, the behaviour for Mimas' eccentricity is in agreement with the results of
\citet{Meyer.Wisdom-2008}. The eccentricity only stops growing when it reaches
a certain equilibrium eccentricity (here value is
higher \SI{0.052}{}, that was also found by
\citet{Meyer.Wisdom-2008}) and the satellites escape of Mimas-Enceladus 3:2
e-Mimas resonance. Although we do not show in this work, the eccentricities of
Mimas and Enceladus should decrease after they went out of the resonance due to
the tidal effects reaching the current values.

These results show that Mimas' eccentricity could be enlarged and hence affect
the width of the corotation resonance. In the
next section we will study this effect on the capture of particles by the corotation resonance.

\section{Capture in Corotation Resonance}
\label{captura}

In the previous section we saw that the eccentricity
of Mimas increases when Mimas and
  Enceladus pass through a 3:2 resonance during the tidal migration
  process. In this section, we will check if this increase in eccentricity will
enable Mimas to capture particles in corotation resonance. To perform this
study, we created a ring with 10\,000 particles in a region where the resonance
should appear when the eccentricity of Mimas increase, as we can see in Figure
(\ref{migracaomimasenceladusmA}{a}).  Thus the particles were uniformly
distributed with semi-major axis between \SI{162409.9}{km} and \SI{162604.8}{km}
and with mean longitude between \SI{0}{\degree} and \SI{360}{\degree}.  All
particles have its other orbital elements fixed with the value \SI{0.01075838}{}
for eccentricity, \SI{0.004053321}{\degree} for inclination,
\SI{342.0739}{\degree} for ascending node, and \SI{326.3412}{\degree} for
longitude of the pericenter. These fixed orbital elements were based on the
orbital elements of Aegaeon after it was migrated toward Saturn and close to a
position where the resonance should appear.  This choice was made to improve
our chances of capture and we are not interested in finding the best orbital
configuration for the capture, but in the migration process as the cause of the
capture in corotation resonance.

For this experiment we integrate the full equations of motion for the four body
model (Saturn, Mimas, Enceladus and a particle) plus equation
(\ref{forcaperturbadora}) acting only in Mimas and Enceladus and representing
  the tide interaction in these bodies. We argue that the homogeneous rings do
not raise tidal bulges in the planet like moons do, since a homogeneous ring
does not raise a tidal bulge. Also, ring's particles have no interaction with
each other by collisions or gravity. We also considered an oblate Saturn for all
particles involved in the integration. The integrations were made using
Gauss-Radau spacings described by \citet{Everhart-1985} with initial time step
of 0.1 day.  We used the same initial conditions for Mimas and Enceladus of
table \ref{initialcondition}. It is important to say that, although all 10\,000
particles were integrated at the same time they do not interact with each other.
Thus it is a four body problem plus Saturn oblateness for each particle.

In panels (d) and (h) of Figure (\ref{migracaomimasenceladusmA}) we can see the
evolution of Mimas' eccentricity and in panels (a), (b), (c), (e), (f) and (g)
the effects this evolution make in the particles of the ring.

In the very beginning, despite of some small variation, Mimas' eccentricity is
not increasing as we can see in Figure
(\ref{migracaomimasenceladusmA}{d}). After some time 
the satellites enter the 3:2 e-Mimas resonance and Mimas'
eccentricity increases affecting the ring, as
shown in Figure (\ref{migracaomimasenceladusmA}{b}).

In Figure (\ref{migracaomimasenceladusmA}{c}) it is possible to see trapped
particles in the resonance which had moved upward
due to Mimas migration. In the panels (e), (f) and (g) of
Figure (\ref{migracaomimasenceladusmA}) we see more explicitly this phenomena
where we clearly see six structures moving outside our initial ring. Those six
lobes are consistent with a 7:6 corotation resonance. It is also possible to see
that some particles were dragged outwards.

To explicitly show that the eccentricity variation is the mechanism responsible
for the capture in corotation resonance, we made other
experiment passing the corotation resonance through
the ring of particles, but with Mimas not varying its
  eccentricity. For that we take off Enceladus of the simulation and initiate
Mimas with its
  present eccentricity of 0.02 and its semi-major axis \SI{100}{km} below the
  value shown in the table (\ref{initialcondition}). For all other orbital elements,
it was maintained the values of table (\ref{initialcondition}), and then it was
applied the migration process to Mimas. The ring of particles was the same as
the last experiment.

\begin{figure}
 \centerline{\includegraphics[scale=0.80,angle=0]{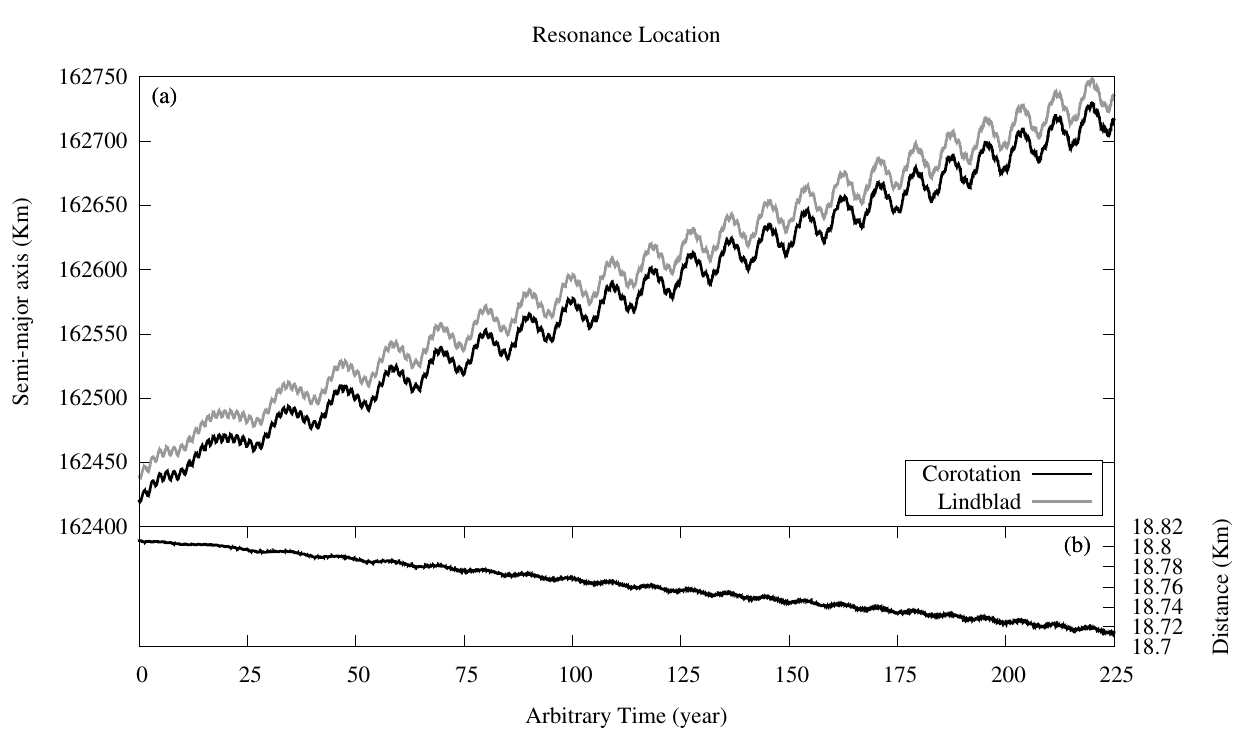}}
 \caption{In panel (a) we present the localization of Corotation and Lindlblad
   resonances. These locations were found following the technique of \citet{Foryta.Sicardy-1996},
   with equations of
   \citet{Renner.Sicardy-2006}. The Corotation and Lindlblad location move
   because Mimas is migrating during this simulation. In (b) it shows the
   distance between these resonances.}
 \label{locdif}
\end{figure}

We noted that the capture of particles doesn't occur due to overlap between the
Lindblad and corotation resonances.  These two resonances have a separation about \SI{19}{km} (Figure (\ref{locdif}{b}) and
the particles in the ring trapped by the corotation resonance feel the Lindblad perturbation
\citep{Moutamid.etal-2014}. These simulations show that, in our accelerated tidal scenario, 
the overlap of resonances isn't the dominant mechanism to capture of particles.

\begin{figure}
 \centerline{\includegraphics[scale=0.82,angle=0]{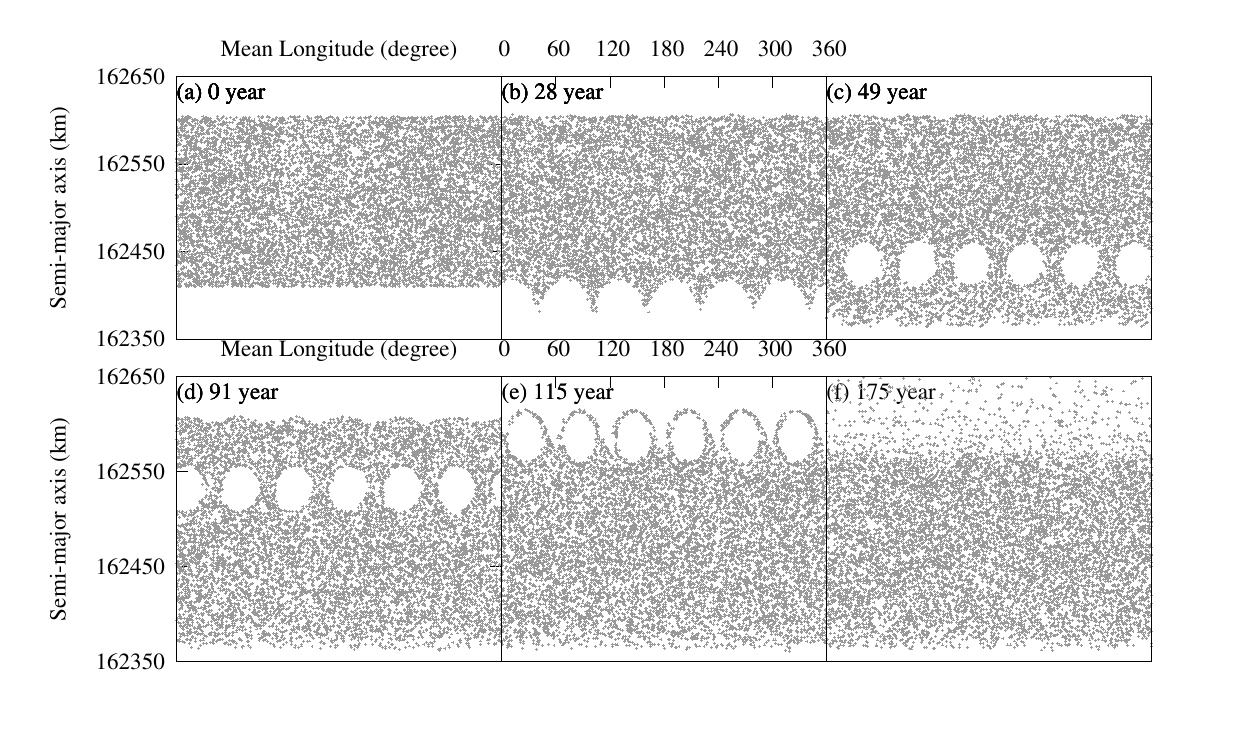}}
 \caption{In these six snapshots shown we had taken off Enceladus of the
   simulation. Without the eccentricity enhancement due to 3:2 resonance there
   is no capture in the corotation resonance. In last snapshot, all corotation
   sites have passed through the ring without capture any particle.}
 \label{multiplotC}
\end{figure}

We can see that holes appear while the corotation resonance passes
  through the ring of particles (see panels (a) to (f) of Figure
  (\ref{multiplotC})). As there are no particles inside these
holes, we can affirm that particles weren't
captured. This effect shows that the corotation resonance cannot capture neither
loose particles unless it changes its width. However, we can see that some
particles stay close to the corotation edge (Figure
(\ref{multiplotC}{e})) and Figure (\ref{multiplotC}{f})). These particles are
temporary captured in a region known as resonance stickiness
\citep{Contopoulos.Harsoula-2010}. These particles move in the border of the
lobes of the corotation resonance for a time and then they escape.

With these results, we show that our scenario could explain the formation of the
arc of the G-ring. In this scenario, we used a migration rate much above of that
generated by tidal effect, and with a realistic migration rate, it must work as
the same way. Actually, we had made our first experiments with rates consistent
with tidal migration and observed the robustness of the corotation resonance.

In one of our first experiments, which we have not shown here, we created an arc
of 1847 particles in the resonance of Mimas where G ring arc was in date January
1$^{st}$, 2004 \citep{Hedman.etal-2010a}, and simulated Mimas migrating with a
velocity equal to \SI{9.0E-9}{km/year} toward Saturn by \SI{1000}{years}. The
final of simulation we found that 98\% of the particles was in corotation with
Mimas, corroborating the robustness of the corotation resonance.

In this scenario, it was needed a high initial eccentricity for Enceladus to
force the escape of Mimas from the 3:2 e-Mimas resonance, but it is not a
problem for our scenario because several other processes could take Mimas out of
this resonance, for instance, an eventual resonance between Mimas and Dione.

\section{Evolution of the Capture Resonance}
\label{evolution}

\begin{figure}
\centerline{\includegraphics[scale=0.3,angle=0]{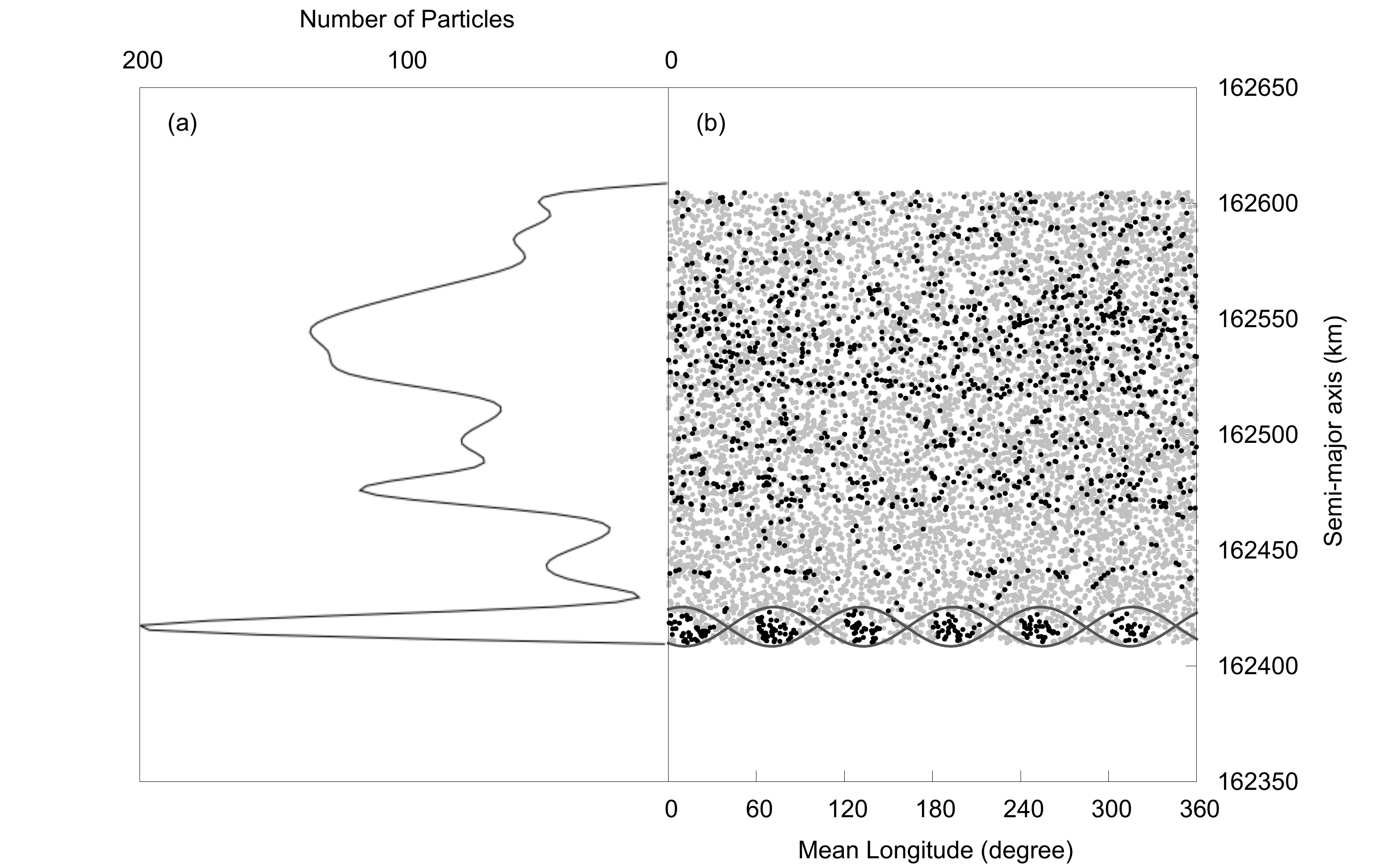}}
\caption{Evolution of the capture in corotation resonance through the ring of
  particles. Panel (b) of this graph shows the initial conditions of the
  captured particles. In panel (a) it was shown a histogram of captured
  particles in ranges of \SI{10}{km} for the semi-major axis.}
 \label{capturadas}
\end{figure}

In the last section, we observed that the particles were captured in corotation
resonance and dragged out of the ring we created. Now we will analyse these
captures through the history of the captures. In Figure (\ref{capturadas}) we
have the ring of particles in time 0 of our simulation and we indicate the
particles that were captured. In the Figure (\ref{capturadas}b), we identified
with grey points the particles which were not captured in the Mimas migration
process and with black points the particles which were captured in this process
and didn't escape the corotation sites during the simulation. The final state of
these captured particles (black points in Figure (\ref{capturadas}b)) can be
seen in panel (g) of Figure (\ref{migracaomimasenceladusmA}) occupying the
corotation sites above the ring of particles.

In the bottom part of the Figure (\ref{capturadas}b), we saw a draft of the 7:6
corotation sites in its initial condition produced by Mimas eccentricity. The
amplitude of the curves obeys the width of the corotation for the initial Mimas
eccentricity (equation (\ref{larguradacorrotacao})). With the Mimas migration,
the width of this curve will enlarge. This enlargement will produce the captures
in the corotation resonances.

\begin{figure}
 \centerline{\includegraphics[scale=0.7,angle=0]{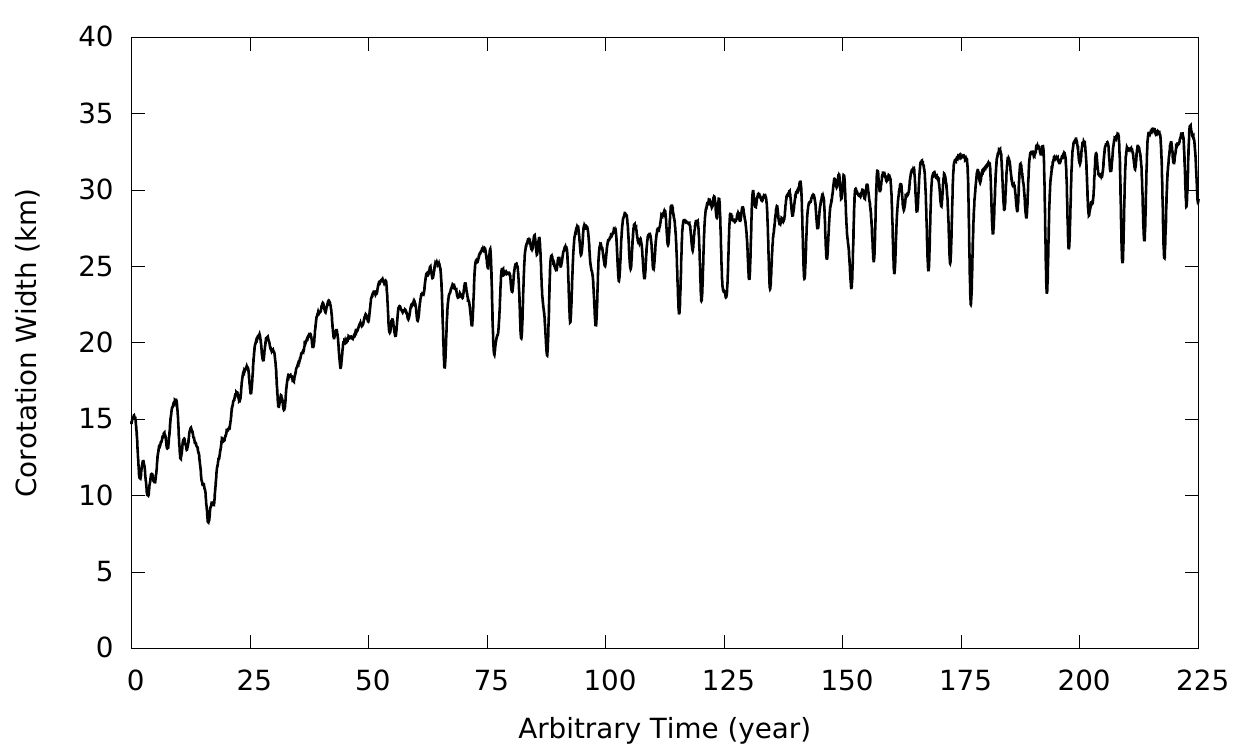}}
 \caption{Amplitude evolution of the corotation resonance width based on
   equation (\ref{larguradacorrotacao}) for each value of the Mimas eccentricity
   of Figure (\ref{migracaomimasenceladusmA}).}
 \label{width}
\end{figure}

We can see in the panel (b) of Figure (\ref{capturadas}) that the black points are
spread in the ring and they show some structures. In panel (a) we show the
number of captured particles in bins of \SI{10}{km} for the semi-major axis. The
excess particles in the low part of the disc occur due to
Mimas initial eccentricity, as the corotation resonance has an initial width showed by the embedded black points encircled by the
draft representing the corotation curves. Observing the structures and the
  histogram of the graph, it suggests that we hadn't a homogeneous rate of
  captures. Thus, the capture is not continuous, but
in steps. The explanation for that
  can be obtained observing the graphs of Figure \ref{migracaomimasenceladusmA}
as the eccentricity oscillates while increases.

The capture by the corotation resonance is complex. When Mimas was captured by
the 3:2 resonance with Enceladus its eccentricity began to increase but not in
a uniform way. In Figure (\ref{width}) we see the evolution of this width based
on equation (\ref{larguradacorrotacao}) and the history of
Mimas eccentricity shown in Figure
(\ref{migracaomimasenceladusmA}). The net result of the process double the
initial width, however locally it increases and decreases in a very noise form.
Thus, particles were captured while there are others that escape from the corotation resonance. 
The particles that were captured and escaped are the ones in the
stickiness caused by the corotation resonance observed in Figure
(\ref{multiplotC}). This complex dynamic explains the structures observed in
Figure (\ref{capturadas}).

This feature can be understood when we consider the capture probability in the
corotation resonance, which depends on perturbing eccentricity
\citep{Quillen-2006}, in our case the perturber is Mimas. The Mimas'
eccentricity increases non-uniformly (Figure \ref{migracaomimasenceladusmA}), so
the capture probability oscillates during the simulation. When the eccentricity
increases, it favours the capture, and when it decreases, the escape
likelihood increases.  The result is a relative capture
probability, because sometimes it is more likely
to capture particles, but sometimes, it is less likely. It was that which creates the structures
observed in Figure (\ref{capturadas}).

These results lead us to conclude that the migration sweeping over the
particles, as shown in Figure (\ref{migracaomimasenceladusmA}), 
with the change in the corotation resonance width producing a change in Mimas' eccentricity, as shown in Figure
(\ref{capturadas}), is the
  process which could explain the capture of the one particle, or a group of particles, which produced the arc of
the G-ring.

\section{Conclusions}
\label{conclusao}

In this paper, we showed that it is possible to explain the formation of the arc
of the G-ring through a plausible scenario where the Saturn tides play the main
role.

The Saturn's tide could vary the semi-major axis of the satellites, which make them
cross several resonances. In this process a low Mimas' eccentricity could have
increased and could have caused the enlarging of the corotation region.

In our experiment, when Mimas created the region of corotation particles, we had
populated all the six structures of the corotation. Then, if there was a ring of particles in the
corotation resonance capture region, we should observe other groups
today. However, there exists only one arc. Our hypothesis is still plausible,
since other effects, such as Poynting-Robertson drag may have destroyed the
other groups, or even the region was not as homogeneous as we created them in
this scenario. We believe that further studies for the arc of the G-ring
formation should solve this problem. Other studies could also answer if the arc
was formed by agglomeration or erosion, and all these information together could
clarify how and why we have only one group observed in 7:6 corotation resonance.

\section*{Acknowledgments}

The authors want to thank Bruno Sicardy for the prolific discussions and helpful
suggestions, the anonymous referee for his helpful suggestions, and the
Brazilian science funding agencies CAPES, CNPq and FAPESP (grant 2011/08171-3).

\bibliographystyle{mn2e}
\bibliography{formingtheGringarc}

\appendix

\section[]{}

\citet{Poulet.Sicardy-2001} affirm that the attraction of the tidal bulge raised
on a planet by a satellite outside the synchronous orbit results in a gain of
angular momentum by the satellite. This causes the orbit of the satellite to
expand and the rate of change of its semi-major axis is
\begin{equation} 
  \label{taxasemieixomario}
  \frac{\dot{a}}{a} = 3 \left( \frac{G}{M_p} \right)^{1/2} \, k_{{}_{2P}}
                               \frac{R_p^5}{Q_P}
                               \frac{m}{a^{13/2}}, 
\end{equation}
where the parameters $a$, $\dot{a}$, $m$, are the semi-major axis, its
variation and the mass of the satellite, whereas $G$ is the gravitational
constant, $M_p$ is the mass of the planet, $k_{{}_{2p}}$ is the Love number of
the planet and $Q_p$ is the dissipation factor.

We consider that the semi-major axis of Mimas and Enceladus
suffered changes described by Equation (\ref{taxasemieixomario}). If we
integrate this equation, keeping $M_p$, $k_{{}_{2p}}$ and
$Q_p$ constant over time and using parameters of Mimas and Enceladus, we can
find the values of semi-major axis for Mimas
which enables its trapping with Enceladus into a resonance in
the remote past.

The first consideration is that the physical properties of the planet are
constant, so we can write that
\begin{equation} 
  \label{constante}
  \xi = 3 \left( \frac{G}{M_p} \right)^{1/2} \, k_{{}_{2P}} \, \frac{R_p^5}{Q_P},
\end{equation}
where $\xi$ is a constant depending on the planet parameters. Consequently, the
equation (\ref{taxasemieixomario}) becomes
\begin{equation} 
  \label{taxasemieixomario1}
  \dot{a} = {a^{-11/2}} \, \xi \, {m}.
\end{equation}
And integrating it we have
\begin{equation} 
  \label{integral}
  \frac{2}{13} \left( {a^{13/2}} - {a_{{}_0}}^{13/2} \right) = \Delta t \, \xi \, m.
\end{equation}
where the integral constant $a_{{}_0}$ is the value of semi-major axis at
$t_{{}_0}$, and $\Delta t=t-t_{{}_0}$. Calculating the value of this integral
with the parameters of Mimas, we have
\begin{equation} 
  \label{variasemieixomaiormimas}
  \frac{2}{13} \left( {a_{{}_M}^{13/2}}-{a_{{}_{0M}}^{13/2}} \right) = 
  \Delta t_{{}_M} \, \xi \, m_{{}_M}, 
\end{equation}
where ${a_{{}_M}}$ and $a_{{}_{0M}}$ are the current and ancient semi-major axis
of Mimas, respectively, ${m_{{}_M}}$ is Mimas' mass and $\Delta t_{{}_M}$ is the
time spent for Mimas to vary its position from
$a_{{}_{0M}}$ to ${a_{{}_M}}$.

Using that integral with parameters of Enceladus, we get
\begin{equation} 
  \label{variasemieixomaiorenceladus}
  \frac{2}{13} \left( a_{{}_E}^{13/2} - a_{{}_{0E}}^{13/2} \right) = 
  \Delta t_{{}_E} \, \xi \, m_{{}_E}, 
\end{equation}
where $M$ was replaced with $E$.

Now, if we consider that $t$ is the current time and $t_{{}_0}$ is an instant
that these satellites would be in resonance we have
$\Delta t_{{}_M} = \Delta t_{{}_E}$, and also the semi-major axes rate of Mimas
and Enceladus for the resonance is $\alpha={a_{{}_{0M}}/a_{{}_{0E}}}$. Then,
from equations (\ref{variasemieixomaiormimas}) and
(\ref{variasemieixomaiorenceladus}), we find that
\begin{equation}
  \label{semieixomaiormimasnopassadoanexo}
  a_{{}_{0M}} = \left[ 
                \frac{\left(\dfrac{m_{{}_E}}{m_{{}_M}}\right) \, a_{{}_M}^{13/2}-a_{{}_E}^{13/2}} 
                     {\left(\dfrac{m_{{}_E}}{m_{{}_M}}\right) - \dfrac{1}{\alpha^{13/2}}}
                \right]^{2/13}. 
\end{equation}
Therefore, with equation (\ref{semieixomaiormimasnopassadoanexo}) we calculate
an approximate value for the ancient semi-major axis of Mimas, just when that
satellite was trapped in resonance with Enceladus, given an appropriate
$\alpha$.

\bsp

\label{lastpage}

\end{document}